\documentclass[journal]{IEEEtran}

\usepackage[T1]{fontenc}
\usepackage[utf8]{inputenc}
\usepackage{amsmath,amsfonts,amssymb}
\usepackage{amsthm}
\usepackage{graphicx}
\usepackage{cite}
\usepackage{xcolor}
\usepackage{mathptmx} 
\usepackage{enumitem} 
\usepackage{tikz}
\usetikzlibrary{positioning, arrows.meta, calc, shadows}
\usepackage{booktabs}
\theoremstyle{plain}
\newtheorem{theorem}{Theorem}
\newtheorem{proposition}{Proposition}
\newtheorem{lemma}{Lemma}
\newtheorem{definition}{Definition}
\newtheorem{axiom}{Axiom}

\begin{document}

\title{Decoupling Correctness from Policy: A Deterministic Causal Structure for Multi-Agent Systems}

\author{Zhiyuan~Ren,~\IEEEmembership{Member,~IEEE},
        Tao~Zhang,
        Wenchi~Cheng,~\IEEEmembership{Senior~Member,~IEEE}
\thanks{This work was supported by the National Key Research and Development Program of China (No. No.2023YFC3011502).}%
\thanks{Zhiyuan Ren, Tao Zhang and Wenchi Cheng are with the School of Telecommunications Engineering, Xidian University, Xi'an 710071, China.}
\thanks{Corresponding author: Zhiyuan Ren (zyren@xidian.edu.cn).}%
}

\maketitle

\begin{abstract}
In distributed multi-agent systems, correctness is often entangled with operational policies such as scheduling, batching, or routing, which makes systems brittle since performance-driven policy evolution may break integrity guarantees. This paper introduces the Deterministic Causal Structure (DCS), a formal foundation that decouples correctness from policy. We develop a minimal axiomatic theory and prove four results: existence and uniqueness, policy-agnostic invariance, observational equivalence, and axiom minimality. These results show that DCS resolves causal ambiguities that value-centric convergence models such as CRDTs cannot address, and that removing any axiom collapses determinism into ambiguity. DCS thus emerges as a boundary principle of asynchronous computation, analogous to CAP and FLP: correctness is preserved only within the expressive power of a join-semilattice. All guarantees are established by axioms and proofs, with only minimal illustrative constructions included to aid intuition. This work establishes correctness as a fixed, policy-agnostic substrate—a "Correctness-as-a-Chassis" paradigm—on which distributed intelligent systems can be built modularly, safely, and evolvably.
\end{abstract}


\section{Introduction}

\IEEEPARstart{A}{ core tension} in designing distributed multi-agent systems (MAS) lies in reconciling the goal of agent \textbf{autonomy} with the need for guaranteed system-wide \textbf{correctness}. This tension often arises from a deep-seated entanglement between an agent's operational \textbf{policy}, which dictates its autonomous behavior, and the system's underlying \textbf{structural correctness} (i.e., the integrity of its causal history). When an agent's behavioral logic is intertwined with the mechanisms that ensure historical integrity, the system becomes brittle: evolving an agent's policy to improve performance or adapt to new conditions risks violating global correctness guarantees. This paper confronts this core MAS challenge directly, focusing on the domain of collaborative, protocol-adherent MAS where the foundational problem is to establish deterministic correctness amidst asynchrony and policy diversity, rather than to defend against malicious, protocol-violating behavior.

This paper asks: can we achieve a formal \textbf{decoupling of correctness from policy}? We propose that a system's correctness should be anchored to a policy-agnostic \textit{structural invariant}---a mathematical object whose integrity is guaranteed regardless of the specific, and possibly non-deterministic, policies executed by the agents. Our answer is a constructive proof centered on a \textbf{Deterministic Causal Structure (DCS)}. We demonstrate that the complete causal history of agent interactions can be proven to converge to a unique, globally consistent \textit{Provenance Directed Acyclic Graph (Provenance DAG)}. This DAG serves as the correctness invariant.

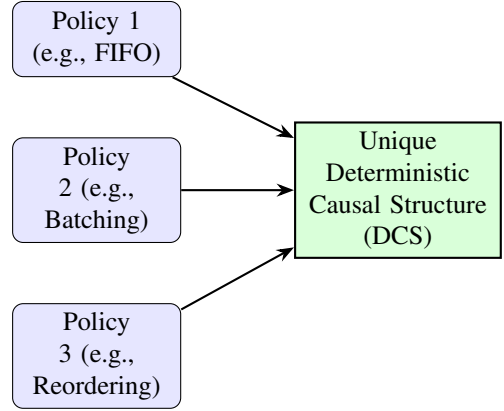
\begin{figure}[h!]
\centering
\begin{tikzpicture}[
    node distance=0.8cm and 1.5cm,
    policy/.style={rectangle, draw, rounded corners, fill=blue!10, text width=2cm, align=center, minimum height=1cm},
    dcs/.style={rectangle, draw, thick, fill=green!15, text width=2.5cm, align=center, minimum height=1cm},
    arrow/.style={- Stealth, thick}
]
    \node[policy] (p1) {Policy 1 (e.g., FIFO)};
    \node[policy] (p2) [below=of p1] {Policy 2 (e.g., Batching)};
    \node[policy] (p3) [below=of p2] {Policy 3 (e.g., Reordering)};

    \node[dcs] (dcs_node) [right=of p2] {Unique \\ Deterministic \\ Causal Structure (DCS)};
    
    \draw[arrow] (p1) -- (dcs_node);
    \draw[arrow] (p2) -- (dcs_node);
    \draw[arrow] (p3) -- (dcs_node);
    
\end{tikzpicture}
\caption{The core principle of Structure-Policy Decoupling. This illustrates how, for the \textbf{exact same set of generated events}, different \textbf{operational policies} (such as message ordering, batching, or reordering) all result in the construction of the identical, unique Deterministic Causal Structure (DCS). The correctness of the causal history is thus decoupled from the infrastructural mechanics of its delivery.}
\label{fig:conceptual_overview}
\end{figure}

It is crucial to define the scope of this decoupling precisely. The policy-agnosticism guaranteed by DCS separates the \textbf{Recording Mechanism} from the \textbf{Execution Path}. The system's final state, which is determined by the specific set of contributions generated over time (the Execution Path), remains fundamentally dependent on the agents' high-level generation policies. The guarantee of DCS is that for any given execution path, the infrastructural mechanism by which that history is recorded is deterministic and unique, regardless of the underlying operational policies (e.g., routing, scheduling) that delivered the information, as shown in Fig.~\ref{fig:conceptual_overview}.

To establish this, we present a complete theory that rests on three core propositions:
\begin{enumerate}
    \item \textbf{Existence and Uniqueness of the DCS (Theorem~\ref{thm:A}):} We prove that under a minimal set of axioms, the global causal history graph (the DCS) exists, is unique up to isomorphism, and is constructively verifiable.
    \item \textbf{The Decoupling Invariant (Theorem~\ref{thm:B}):} We prove that this DCS is a policy-agnostic invariant. For any two admissible policies (e.g., different schedulers or routing strategies), the resulting DCS graphs are guaranteed to be isomorphic. This formally separates the system's correctness from its operational policies.
    \item \textbf{Observational Equivalence (Proposition~\ref{prop:C}):} We prove that for a broad class of applications, two executions are observationally indistinguishable \textit{if and only if} their DCS graphs are isomorphic. This establishes the DCS as the minimal and sufficient carrier of information for correctness.
\end{enumerate}

This decoupling provides a new paradigm for system design. Instead of reasoning about complex, holistic system behaviors, developers can first establish the correctness of the structural invariant, and then safely and independently explore the vast space of performance-optimizing policies. Our core contributions are:
\begin{enumerate}
    \item We formalize the structure-policy entanglement problem and propose its decoupling as a core design principle for robust distributed systems.
    \item We present the theory of a Deterministic Causal Structure (DCS) and prove it serves as a policy-agnostic invariant for correctness (Theorems~\ref{thm:A} \& \ref{thm:B}).
    \item We complete the theory by proving that DCS isomorphism is a necessary and sufficient condition for observational equivalence for a wide range of applications (Proposition~\ref{prop:C}).
    \item We formally prove that the structural guarantee of a DCS is strictly stronger than the value convergence of models like CRDTs.
\end{enumerate}

The remainder of this paper is organized as follows. Section II reviews related work to position our contribution. Section III establishes our formal model and core axioms. Section IV presents the core theoretical results of the DCS theory. Section V provides a set of illustrative constructions to intuitively demonstrate the implications of our theoretical results. Section VI discusses the broader implications and theoretical boundaries of our work. Finally, Section VII concludes the paper.

\section{Related Work}

The DCS theory is situated at the intersection of research in distributed systems and multi-agent systems. To clearly position the uniqueness and contributions of our work, this section provides a systematic comparison of our theory against several key related research areas. We will sequentially discuss: classical distributed consensus protocols; replicated data types and eventual consistency models; causal consistency theories; DAG-based BFT consensus; and data provenance. Through these comparisons, we will elucidate the fundamental innovation of DCS in guaranteeing structural determinism.

\subsection{Distributed Consensus and Total Order Broadcast}

Classical consensus protocols, foundational to distributed systems, aim to solve the State Machine Replication (SMR) problem by enforcing a \textbf{Total Order Broadcast} \cite{lamport1982byzantine}. Early practical solutions like Viewstamped Replication \cite{okiliskov1988}, and later the widely-known Paxos \cite{lamport1998part} and Raft \cite{ongaro2014search}, provided mechanisms to achieve this. However, the performance limitations of this strictly sequential execution on modern multi-core architectures are well-documented \cite{wojciechowski_state-machine_2017, burgos_exploiting_2022}. Consequently, a significant body of recent research has focused on overcoming this bottleneck, through avenues like \textit{parallel state machine replication} \cite{burgos_exploiting_2022, wojciechowski_state-machine_2017}, hybrid systems that switch between deterministic and optimistic execution \cite{kobus_hybrid_2018}, and state partitioning to enable scalability \cite{nogueira_elastic_2017}.

However, a common thread unites these protocols: they achieve correctness by definitionally \textbf{entangling it with a specific ordering policy}. To linearize naturally concurrent events, the system must employ a policy (e.g., a leader's sequencing decision) to dictate an arbitrary order between them \cite{nogueira_elastic_2017}. The final log---the carrier of correctness---is therefore a direct artifact of this policy. The DCS theory offers a fundamentally different approach based on decoupling. Instead of enforcing a policy-dependent total order, it seeks consensus on the objective, policy-\textit{independent} \textbf{Causal Partial Order}. The DCS preserves the natural concurrency of interactions, capturing the "ground truth" of what happened without imposing an artificial timeline, thus providing a stable structural invariant upon which diverse performance strategies can be independently optimized.

\subsection{Replicated Data Types and Value Convergence}

Conflict-free Replicated Data Types (CRDTs) \cite{shapiro2011conflict, shapiro2011convergent} are philosophically the closest to our work, as they also embrace concurrency. While earlier techniques like Operational Transformation (OT) \cite{ellis1989concurrency} addressed concurrent updates, they often required central coordination, a challenge that persists in modern implementations \cite{gadea_control_2020}. CRDTs, grounded in theoretical principles like the CALM theorem \cite{hellerstein2010calm}, instead guarantee that replicas converge to the same final \textbf{value} in a decentralized manner, regardless of message delivery order. This powerful form of \textbf{Value Convergence} has seen significant optimization, such as Delta State CRDTs which reduce network overhead by propagating only incremental changes \cite{almeida2016delta}.

However, this guarantee of value convergence is achieved by wedding correctness to a policy of \textit{causal indifference}. The semilattice merge function is, by design, insensitive to the causal relationship between operations; it only cares about the set of operations to be merged. This focus on value at the expense of causality has been a subject of further research \cite{zarzour_b-set_2012, kokocinski_mixing_2022}, and has led some systems to use a partially-ordered log, such as a DAG, as the underlying structure for the historical record, while relying on CRDTs at the application layer to interpret the state from this log \cite{karlsson_vegvisir_2018, auvolat_merkle_2019}. This leads to what we have termed \textit{Structural Ambiguity}: as formally proven in Proposition~\ref{prop:crdt_separation} and illustrated in Fig.~\ref{fig:fig6}, two executions with non-isomorphic causal histories can produce the same final value. The DCS theory provides a strictly stronger guarantee by establishing a policy-agnostic \textbf{structural invariant}, resolving the structural ambiguity that CRDTs ignore.

\subsection{Causal Consistency and Logical Clocks}

Causal consistency, often implemented with tools like Lamport Clocks \cite{lamport1978time} and Vector Clocks \cite{fidge1988timestamps, mattern1989virtual}, ensures that events are processed in an order that respects causality. This model is often offered as an intermediate guarantee between strong consistency and eventual consistency \cite{al-ekram_multi-consistency_2010}, ensuring that causally related operations are seen by all replicas in the same order. This model found significant practical application in large-scale systems like Amazon's Dynamo, which used vector clocks to manage versioning in a highly available key-value store \cite{decaneto2007dynamo}.

However, this guarantee is fundamentally a \textbf{local reasoning tool}, not a global, convergent invariant. While it enforces a correct \textit{local processing policy}, it does not guarantee that the resulting global history graph will be unique across all nodes. Since concurrent, causally independent operations are not ordered with respect to each other, different replicas may process them in different sequences, resulting in divergent and non-isomorphic views of the history that are both, from their local perspectives, causally consistent \cite{al-ekram_multi-consistency_2010}. DCS solves this by shifting the burden of correctness from the local processing policy to the \textbf{data structure itself}. By enforcing immutable, universally consistent metadata (\texttt{rid}, \texttt{parents}), the DCS theory guarantees the emergence of a single, global structural invariant, decoupling the system's ground truth from the partial, policy-dependent views of individual agents.

\subsection{DAG-based BFT Consensus}

Recent high-performance BFT consensus protocols leverage a Directed Acyclic Graph (DAG) structure to decouple data dissemination from ordering. Systems like Hedera Hashgraph \cite{baird2016swirlds}, Aleph \cite{gizinski2022aleph}, Narwhal and Tusk \cite{danezis2021narwhal}, and Bullshark \cite{giridharan2022bullshark} have demonstrated significant performance gains using this approach. In these systems, nodes first form a DAG of transactions, which efficiently records the causal partial order. However, in all these designs, the DAG is merely a \textbf{means to an end}: the ultimate goal remains to establish a \textbf{Total Order}. This is evident in recent research, where DAGs are used to enhance existing protocols but must still be resolved into a final, linear chain \cite{jo_enhancing_2024}, and where significant effort is focused on optimizing the secondary \textbf{ordering policy} itself, using everything from lightweight broadcast primitives \cite{dai_lightdag_2024, zhou_plaindag_2025} and theoretical quorum analysis \cite{ladelsky_quorum_2025} to learning-based approaches with GNNs \cite{diallo_optimized_2025}.

Correctness is thus defined by the output of this policy-dependent linearization process. In sharp contrast, the DCS theory posits that the unique, globally consistent DAG is \textbf{the end itself}. We provide a \textbf{structural consensus}, not an ordering consensus. The goal is not to flatten the rich partial order into a policy-dependent total order, but to agree on the policy-\textit{independent} structure of the partial order itself.

\subsection{Data Provenance and Scientific Workflows}

Our work is deeply connected to Data Provenance, as the DCS is, by definition, a Provenance DAG. The field of data provenance, codified in standards like the W3C PROV model, studies the origin and history of data, with significant research on representing and querying these graphs \cite{cheney2009provenance}. However, traditional provenance systems often focus on representation, implicitly \textbf{entangling the integrity of the graph with a specific recording policy}, such as a trusted, centralized logger. This assumption breaks down in decentralized environments, leading recent research to turn to distributed ledger technology (DLT) to construct a secure and tamper-evident data lineage \cite{nepal_secure_2024, nepal_secure_2024-1}, sometimes requiring novel consensus mechanisms tailored for provenance to ensure the history is trustworthy \cite{miller_prism_2023}.

The DCS theory addresses this same fundamental problem—how to construct this graph in a decentralized, trustless environment—but from a different perspective. Instead of relying on an external, application-level consensus protocol or a generic DLT, it achieves the \textbf{decoupling of the graph's integrity from any recording policy} through its axiomatic foundation. Correctness is not guaranteed by an external mechanism layered on top, but by the axiomatic, immutable properties of the data contributions themselves. The DCS is therefore not a new model for representing provenance, but a \textit{consensus protocol for constructing a policy-agnostic, verifiable Provenance DAG}. It provides the missing formal foundation for building truly decentralized and auditable scientific workflows.

\subsection{Blockchain and Distributed Ledger Technologies}

Classical blockchains, exemplified by Bitcoin \cite{nakamoto2008bitcoin} and Ethereum \cite{buterin2014ethereum}, establish an immutable ledger by achieving a probabilistic \textbf{Total Order Consensus}. This model's correctness is inextricably \textbf{entangled with a resource-intensive admission policy}, like Proof-of-Work (PoW) or Proof-of-Stake (PoS) \cite{king2012ppcoin}, which creates a bottleneck. While DAG-based blockchains were developed to overcome these limitations by allowing concurrent block creation \cite{wang_understanding_2020, yang_codag_2019}, simply adopting a DAG structure does not eliminate the need for higher-level policies. Advanced systems like MorphDAG and RT-DAG, for instance, superimpose complex workload-aware or real-time ordering policies on the DAG to manage concurrency and transaction patterns \cite{zhang_morphdag_2024, liao_rt-dag_2024}.

The DCS paradigm offers a fundamentally different vision for a distributed ledger. It not only decouples the ledger's integrity from a resource-intensive \textit{admission policy} like PoW, but also from the complex \textit{workload-aware or real-time ordering policies} required by advanced DAG blockchains. Instead of a linear chain, the ledger \textit{is} the unique Provenance DAG itself. Correctness is not defined by a single, policy-ordered history, but by the policy-agnostic, intrinsic structure of all interactions. This suggests a new type of distributed ledger technology where the primary goal is not total ordering, but consensus on the complete, causally-rich, and policy-independent graph of the interaction history.

\begin{table*}[h!]
\centering
\caption{Conceptual Comparison of DCS with Related Distributed Systems Paradigms}
\label{tab:related_work_summary}
\resizebox{\textwidth}{!}{%
\begin{tabular}{@{}l|llll@{}}
\toprule
\textbf{Paradigm} & \textbf{Core Goal} & \textbf{Object of Consensus} & \textbf{Structure} & \textbf{Policy Dependence} \\ \midrule \midrule

\textbf{DCS (Our Work)} & 
\begin{tabular}[c]{@{}l@{}}\textbf{Structural Correctness} \\ (Decoupled from Policy)\end{tabular} & 
\begin{tabular}[c]{@{}l@{}}\textbf{Causal Partial Order} \\ (The unique DAG itself)\end{tabular} & 
\begin{tabular}[c]{@{}l@{}}Deterministic \\ Causal DAG\end{tabular} & 
\textbf{Policy-Agnostic} \\ \midrule

\begin{tabular}[c]{@{}l@{}}Classical Consensus \\ (e.g., Paxos, Raft)\end{tabular} & 
\begin{tabular}[c]{@{}l@{}}State Machine Replication \\ (SMR)\end{tabular} & 
Total Order Broadcast & 
Linear Log / Chain & 
\begin{tabular}[c]{@{}l@{}}Tightly Entangled \\ (e.g., Leader's sequence)\end{tabular} \\ \midrule

\begin{tabular}[c]{@{}l@{}}Replicated Data Types \\ (CRDTs)\end{tabular} & 
Value Convergence & 
\begin{tabular}[c]{@{}l@{}}Final State Value \\ (Merge Function)\end{tabular} & 
\begin{tabular}[c]{@{}l@{}}Structurally Ambiguous \\ (History is lost)\end{tabular} & 
\begin{tabular}[c]{@{}l@{}}Entangled with \\ Causal Indifference\end{tabular} \\ \midrule

\begin{tabular}[c]{@{}l@{}}Causal Consistency \\ (e.g., Vector Clocks)\end{tabular} & 
Local Causal Ordering & 
\begin{tabular}[c]{@{}l@{}}None (Local validation, \\ no global agreement)\end{tabular} & 
\begin{tabular}[c]{@{}l@{}}Divergent Local Views \\ (Non-isomorphic)\end{tabular} & 
\begin{tabular}[c]{@{}l@{}}Entangled with \\ Local Reception Order\end{tabular} \\ \midrule

\begin{tabular}[c]{@{}l@{}}DAG-based BFT \\ (e.g., Narwhal, Tusk)\end{tabular} & 
High-Throughput SMR & 
\begin{tabular}[c]{@{}l@{}}Total Order \\ (Extracted from DAG)\end{tabular} & 
\begin{tabular}[c]{@{}l@{}}Intermediate DAG, \\ Final Linear Log\end{tabular} & 
\begin{tabular}[c]{@{}l@{}}Tightly Entangled \\ (Secondary ordering protocol)\end{tabular} \\ \midrule

\begin{tabular}[c]{@{}l@{}}Data Provenance \& \\ Blockchain\end{tabular} & 
\begin{tabular}[c]{@{}l@{}}Verifiable History / \\ Asset Transfer\end{tabular} & 
\begin{tabular}[c]{@{}l@{}}Total Order (Blockchain) / \\ None (Traditional Provenance)\end{tabular} & 
\begin{tabular}[c]{@{}l@{}}Linear Chain / \\ Centrally Recorded Graph\end{tabular} & 
\begin{tabular}[c]{@{}l@{}}Entangled with Recording \\ or Admission Policy (e.g., PoW)\end{tabular} \\ \bottomrule
\end{tabular}%
}
\end{table*}

In summary, this review reveals that existing paradigms consistently entangle system correctness with operational policy, a conceptual overview of which is presented in Table~\ref{tab:related_work_summary}. This pervasive entanglement motivates our work to formally decouple structural integrity from policy, which the following sections now develop.

\section{The Formal Framework: Model and Axioms}

To deduce the intrinsic structural laws of multi-agent interactions from first principles, we must first construct a precise formal framework. This section establishes this framework by first defining our system model and fault assumptions, then introducing a minimal interaction model, and finally proposing a set of core axioms. These axioms represent the minimal set of formal constraints required to guarantee the emergence of a Deterministic Causal Structure.

\subsection{System Model and Fault Assumptions}

We consider a standard distributed system composed of a set of agents $V$. The system operates under the following assumptions:
\begin{itemize}[leftmargin=*]
    \item \textbf{Asynchronous Model}: The system is asynchronous. There is no global clock, and we make no assumptions about the relative processing speeds of agents or message delivery times, other than that they are finite.

    \item \textbf{Non-Byzantine Fault Model}: Agents are assumed to be "honest-but-fallible." They correctly follow the protocol rules (i.e., the axioms) at all times, but may fail at any time by crashing (the crash-stop model). This is a deliberate and fitting model for collaborative MAS, where agents are code-driven and operate in a permissioned environment. Unlike open, adversarial systems, the fundamental problem here is not malice, but ensuring resilience and deterministic outcomes despite software faults and network uncertainty. We therefore do not consider arbitrary or protocol-violating (Byzantine) behavior.
    \item \textbf{Unreliable Network Model}: The network connecting the agents is unreliable. It may lose, duplicate, or reorder messages at will. The only liveness guarantee on communication is a weak fairness property, which will be formally stated in Axiom~\ref{ax:fairness}.
\end{itemize}

\subsection{Interaction Model}

The model aims to capture the core process of agents building a shared body of knowledge through atomic contributions, given the system model defined above.

\paragraph{Basic Components} The system consists of a finite \textbf{Set of Agents}, denoted as $V = \{1, \dots, n\}$, and an abstract \textbf{Key Space}, denoted as $K$, used to logically partition interactions.

\paragraph{State and Merging} For any agent $i \in V$ and key $k \in K$, its \textbf{Local State} is denoted as $M_i(k)$. Each key's state belongs to a \textbf{State Space} $(L_k, \sqsubseteq, \sqcup)$, which we require to be a \textbf{directed-complete join-semilattice} (dcpo-join-semilattice). A \textbf{join-semilattice} is a partially ordered set where the join operation $\sqcup$ is associative, commutative, and idempotent.

\paragraph{The Canonical Representation of Interaction: The Contribution}
All interactions in our model are captured by a single, canonical data structure called a \textbf{Contribution}, denoted by $\delta$. We formally define it as a tuple:
\begin{definition}[Contribution]\label{def:contribution}
A contribution $\delta$ is a tuple $(\texttt{rid}, \texttt{parents}, \texttt{payload}, k)$, where:
\begin{itemize}[leftmargin=*, itemsep=0.5ex]
    \item $\texttt{rid} \in \mathcal{R}$ is a globally unique identifier for the contribution.
    \item $\texttt{parents}$ is a finite set of \texttt{rid}s, $\{p_1, p_2, \dots, p_m\}$, representing the direct causal predecessors of this contribution.
    \item $\texttt{payload}$ contains the arbitrary, application-specific data.
    \item $k \in K$ is a key used for logical partitioning of contributions.
\end{itemize}
\end{definition}

\begin{figure}[ht]
\centering
\includegraphics{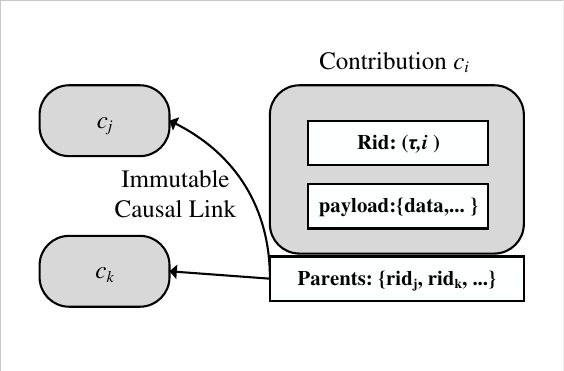}
\caption{The internal structure of a Contribution. The immutable causal history is "soldered" into the data structure itself via the \texttt{parents} field, which contains the unique RIDs of its direct predecessors. This forms an unbreakable, verifiable causal chain.}
\label{fig:contribution_structure}
\end{figure}

\noindent This specific tuple structure is not arbitrary, but is itself a minimal requirement for achieving a DCS. The \texttt{rid} provides a unique identity for each event. The \texttt{parents} set is essential for explicitly encoding the causal dependencies that form the very fabric of the history graph. The separation of \texttt{payload} from the immutable metadata (\texttt{rid}, \texttt{parents}) is the key to enabling policy-agnosticism, as our axioms will only constrain the metadata, leaving the content entirely free. Therefore, this data format is the necessary substrate upon which our axioms operate to guarantee structural convergence.

\paragraph{System Execution Model} We define the \textbf{Universe of Contributions} as the set $\mathcal{R}$ of all possible \texttt{rid}s. For any key $k \in K$, a \textbf{Relevant Agent Set}, $\mathrm{Rel}(k) \subseteq V$, represents the members that subscribe to information for this key. For any agent $i \in \mathrm{Rel}(k)$, its \textbf{Final Mergeable Set}, $\mathsf{Merge}_i(k) \subseteq \mathcal{R}$, is the set of \texttt{rid}s it is guaranteed to eventually receive and merge.

\subsection{Core Axioms}

Having defined the components of our system, we now establish the fundamental rules governing their interaction. The following axioms are not arbitrary postulates, but rather the logical consequences derived from the requirements for achieving a Deterministic Causal Structure in an asynchronous, decentralized environment. We will present them constructively by addressing the key challenges of communication, state management, and historical integrity.

\paragraph{The Communication Postulate}
The first foundational challenge is communication. In an asynchronous, unreliable network, messages can be lost. If any agent is permanently unable to see a piece of the interaction history that others have seen, then no global convergence of any kind is possible. Therefore, any viable theory must begin with a baseline liveness guarantee. We do not need to assume a strong property like reliable or ordered delivery, but we must posit a minimal fairness condition that ensures all relevant information is eventually propagated. This leads to our first axiom.

\begin{axiom}\label{ax:fairness}
\textbf{(Localized Weak Fairness)} For any key $k \in K$, if a contribution $\delta$ belonging to that key is persistently sent, then $\delta$ will eventually be delivered at least once to all agents in the relevant set $\mathrm{Rel}(k)$.
\end{axiom}

\paragraph{The Algebraic Foundation for State}
The second challenge stems from the consequences of asynchronicity: message reordering and duplication. If the result of integrating new information into an agent's local state depends on the order of arrival, the system's outcome will be non-deterministic. To eliminate this ambiguity, the state update mechanism must be inherently order-agnostic. This requires the state space to possess specific algebraic properties—namely, that the merge operation is associative, commutative, and idempotent. A join-semilattice is the formal structure that precisely captures these properties. The requirement of being "directed-complete" is a further technical condition to rigorously handle the limits of infinite executions.

\begin{axiom}\label{ax:semilattice}
\textbf{(Directed-Complete Join Semilattice)} For any key $k \in K$, its state space $(L_k, \sqsubseteq, \sqcup)$ must form a directed-complete \textbf{join-semilattice}.
\end{axiom}

\paragraph{The Integrity of Historical Facts}
Finally, and most critically, even with guaranteed communication and an order-agnostic state model, a DCS is impossible if the historical records themselves—the contributions—are not well-behaved. To construct a single, unambiguous history, each event or "fact" within that history must be immutable and causally well-formed. We enforce this through a set of integrity axioms. First, each fact must have a unique identity and unchangeable content.

\begin{axiom}\label{ax:uniqueness}
\textbf{(Contribution as an Immutable Fact)} Every contribution $\delta$ represents a self-contained, immutable historical fact. To guarantee this, its identity, the \texttt{rid}, must be globally unique, and its substance, the \texttt{payload}, must be immutable upon creation.
\end{axiom}

\begin{axiom}\label{ax:parents}
\textbf{(Immutability of Causal Linkage)} To construct a single, unambiguous causal graph, the relational links between contributions must be permanently fixed. Therefore, the \texttt{parents} set of every contribution $\delta$, which encodes its direct causal dependencies, must be immutable upon creation.
\end{axiom}

Second, beyond immutability, each fact must respect the arrow of time. An event cannot be caused by something that has not yet happened. This principle of causal well-formedness is essential to prevent paradoxes and guarantee an acyclic history.

\begin{axiom}\label{ax:causal}
\textbf{(Causal Well-Formedness)} When an agent generates a new contribution $\delta$, all \texttt{rid}s in its \texttt{parents} set must be drawn from the set of contributions already observed by that agent at the time of creation.
\end{axiom}

\section{Core Theoretical Results}

Having established the formal framework and the core problem of policy-correctness entanglement, this section presents our solution: the formal theory of a \textbf{Deterministic Causal Structure (DCS)} as a policy-agnostic invariant. We will first prove the existence and uniqueness of the DCS (Theorem~\ref{thm:A}), then prove its invariance under any admissible policy (Theorem~\ref{thm:B}), and finally, formally connect this structural isomorphism to the concept of observational equivalence (Proposition~\ref{prop:C}), thus completing our decoupling argument.

\subsection{Theorem A: Existence, Uniqueness, and Constructibility of the DCS}

We begin by proving that in any system adhering to our axioms, the seemingly chaotic interactions will inevitably converge to a single, well-defined mathematical object.

\begin{theorem}[Existence, Uniqueness, and Constructibility of the DCS]\label{thm:A}
Let the global interaction graph be a \textbf{Provenance DAG}, denoted by $G^* = (\mathcal{R}, E)$, where $\mathcal{R}$ is the set of all contribution \texttt{rid}s, and $E = \{ (p \to r) \mid p \in \mathrm{parents}(r) \}$ is the set of causal dependency edges. Under Axioms~\ref{ax:fairness} through~\ref{ax:causal}:
\begin{enumerate}
    \item \textbf{(Existence)} The global interaction graph $G^*$ is a well-defined Directed Acyclic Graph.
    \item \textbf{(Uniqueness)} The structure of $G^*$ (up to graph isomorphism) is unique, independent of the order of contribution generation and propagation.
    \item \textbf{(Constructibility)} For any key $k$ and any agent $i \in \mathrm{Rel}(k)$, its local view of the graph converges, and the limit of its local state $M_i^*(k)$ exists, is unique, and is equal to the join of all payloads in its final mergeable set.
\end{enumerate}
\end{theorem}

\begin{IEEEproof}[Proof Sketch]
The proof demonstrates how distinct subsets of our axioms guarantee each property of the DCS. \textbf{(1) Existence} of a well-defined DAG is guaranteed primarily by Axiom~\ref{ax:causal} (Causal Well-Formedness), which prohibits cycles. \textbf{(2) Uniqueness} of this structure up to isomorphism is a direct consequence of Axioms~\ref{ax:uniqueness} and~\ref{ax:parents}, which enforce the immutability of the graph's metadata (\texttt{rid} and \texttt{parents}). Finally, \textbf{(3) Constructibility} of a deterministic state from this unique structure is made possible by Axioms~\ref{ax:fairness} and~\ref{ax:semilattice}, which ensure that all information is eventually received and can be merged in a policy-agnostic, order-independent manner. The full formal proof is detailed in Appendix~\ref{sec:appendix_theorems}.
\end{IEEEproof}

\subsection{Theorem B: The Decoupling Invariant --- Policy-Agnosticism}

Theorem A establishes the DCS as a stable, unique structure. The next theorem reveals its most powerful property: this structural integrity is completely independent of the agents' behavior, providing the formal basis for our decoupling claim.

\begin{theorem}[The Decoupling Invariant: Policy-Agnosticism]\label{thm:B}
For any two admissible agent policies $P_1$ and $P_2$, which may differ in any aspect of their operation (e.g., scheduling, batching, routing), if they generate the same set of contributions, the resulting global DCS graphs $G_1^*$ and $G_2^*$ are isomorphic.
\end{theorem}

\begin{IEEEproof}[Proof Sketch]
The proof follows directly from the logic of Theorem~\ref{thm:A}. A review of that proof reveals that its every step relies solely on the formal properties of the contribution's metadata (\texttt{rid}, \texttt{parents}), as constrained by our axioms. The proof logic never inspects the \texttt{payload} content, nor does it impose any constraints on the timing, frequency, or ordering of contribution generation, which constitute the agent's policy. Therefore, the structural conclusion of Theorem~\ref{thm:A} is necessarily independent of any specific policy, establishing the DCS as a policy-agnostic invariant. The full proof is detailed in Appendix~\ref{sec:appendix_theorems}.
\end{IEEEproof}


\subsection{Proposition C: Observational Equivalence}

Theorems A and B show we can decouple a structural invariant from policy. But why is this specific invariant the \textit{right} one? This final proposition closes the loop by proving that the DCS structure is precisely the information needed for correctness: two systems are indistinguishable from the outside if and only if their internal structures are the same.

\begin{proposition}[Observational Equivalence]\label{prop:C}
For all upper-layer computations that use only (i) ancestor and concurrency queries on the DCS and (ii) semilattice-homomorphic aggregates over payloads, two executions are observationally indistinguishable if and only if their DCS graphs are isomorphic.
\end{proposition}

\begin{IEEEproof}[Proof Sketch]
The proof proceeds in two directions.
($\Leftarrow$) If the DCS graphs are isomorphic, any structural query (like `is\_ancestor(p,r)`) will yield identical results. Since payloads are immutable and uniquely identified by \texttt{rid}s, any semilattice-based aggregation over them will also be identical. Thus, the executions are observationally indistinguishable.
($\Rightarrow$) If the DCS graphs are non-isomorphic, there must exist a structural difference (e.g., an edge $(p \to r)$ exists in one but not the other). We can then construct a simple structural query that distinguishes the two executions. This proves that the DCS contains the minimal and sufficient information for this class of applications.The full formal proof is detailed in Appendix~\ref{sec:appendix_propositions}.
\end{IEEEproof}

This completes the core of our decoupling theory. The following subsections further solidify its foundations by proving the necessity of our axioms and formally positioning it against existing models.

\subsection{The Theoretical Boundary: Minimality of the Axioms}

The preceding theorems have demonstrated the power of the DCS theory—guaranteeing structural determinism even for fully autonomous ``black box'' agents. A natural and rigorous question to ask is whether we have introduced redundant or overly strong axioms to achieve this powerful guarantee, thereby unnecessarily narrowing the theory's scope of applicability. A complete theory must demonstrate not only the sufficiency of its premises, but also their necessity. The following theorem addresses this question directly by proving the minimality of our axiom set, thereby establishing the sharp and solid boundaries of our theory.

\begin{theorem}[Axiom Minimality]
\label{thm:minimality}
The axiom set \{Axiom~\ref{ax:fairness}, \ref{ax:semilattice}, \ref{ax:uniqueness}, \ref{ax:parents}, \ref{ax:causal}\} is the minimal set required to guarantee a DCS. If any one of these axioms is removed, a constructive counterexample exists where the guarantees of Theorem~\ref{thm:A} fail.
\end{theorem}

\begin{IEEEproof}[Proof Sketch]
The sufficiency of this axiom set is established by the proof of Theorem~\ref{thm:A}. The necessity is proven by constructing counterexamples that demonstrate how removing axioms from each subset shatters a distinct aspect of the DCS guarantee. As detailed in Appendix~\ref{sec:appendix_minimality}, removing axioms from the \textbf{Structural Integrity} set (Axioms 3-5) leads to non-unique or ill-defined causal graphs (Fig.~\ref{fig:fig3}, \ref{fig:fig4}, \ref{fig:fig5}), while removing axioms from the \textbf{Deterministic State Interpretation} set (Axioms 1-2) makes it impossible to derive a consistent, policy-agnostic state from the graph, even if the structure itself were unique (Fig.~\ref{fig:fig1}, \ref{fig:fig2}).
\begin{enumerate}[label=(\roman*), wide, labelindent=0pt, leftmargin=*, itemsep=0.5ex]

    \item \textbf{Removing Axiom~\ref{ax:fairness} (Weak Fairness):} This allows for network partitions where some agents never receive critical contributions, leading to permanently divergent final states. This failure of value convergence is illustrated in Fig.~\ref{fig:fig1}.
    
    \item \textbf{Removing Axiom~\ref{ax:semilattice} (Join-Semilattice):} Without the algebraic properties of a join-semilattice, the merge operation may not be commutative or idempotent. This makes the final state dependent on the arbitrary arrival order of messages, as shown in Fig.~\ref{fig:fig2}.
    
    \item \textbf{Removing Axiom~\ref{ax:uniqueness} (Contribution as an Immutable Fact):} If contribution metadata like the \texttt{rid} is not unique, the global graph becomes ill-defined, as different events may claim the same identity. This is illustrated in Fig.~\ref{fig:fig3}.
    
    \item \textbf{Removing Axiom~\ref{ax:parents} (Immutability of Causal Linkage):} If contribution metadata is not immutable, different executions can result in non-isomorphic but equally valid causal graphs for the same set of events, destroying structural uniqueness, as depicted in Fig.~\ref{fig:fig4}.
    
    \item \textbf{Removing Axiom~\ref{ax:causal} (Causal Well-Formedness):} This permits the formation of forward-references, allowing contributions to form cycles. The global history is consequently no longer a DAG, and the foundational premise of the theory is broken, as shown in Fig.~\ref{fig:fig5}.
    
\end{enumerate}
\end{IEEEproof}

\begin{figure}[htbp]
\centering
\begin{tikzpicture}[
    node distance=1.2cm and 1.8cm, 
    agent/.style={rectangle, draw, thick, minimum size=1cm, font=\bfseries},
    contrib/.style={circle, draw, minimum size=0.8cm},
    state/.style={rectangle, draw, dashed, fill=gray!10, text centered}
]
    \pgfdeclarelayer{background}
    \pgfsetlayers{background,main}

    \begin{pgfonlayer}{background}
        \node[agent] (u) {Agent u};
        \node[agent] (v) [right=of u] {Agent v};
        \node[agent] (w) [below left=of u] {Agent w};

        \node[contrib] (delta) [below right=of w] {$\delta_r$};

        \node[state] (state_u) [below=2.5cm of u] {$M_u^*(k) = a_r$};
        \node[state] (state_v) [below=2.5cm of v] {$M_v^*(k) = \bot$};

        \node at ($(state_u)!0.5!(state_v)$) {\huge $\neq$};
    \end{pgfonlayer}

    \draw[-{Stealth}] (w) -- (delta) node[midway, above left] {creates};
    
    \draw[-{Stealth}, green!60!black, thick] (delta) to[bend left] node[pos=0.7, above, sloped] {delivered} (u);
    \draw[-{Stealth}, red, thick, dashed] (delta) to[bend right] node[pos=0.7, below, sloped] {undelivered} (v);

\end{tikzpicture}
\caption{Violation of Axiom~\ref{ax:fairness} (Weak Fairness). The failure to deliver a contribution to all relevant agents results in permanently divergent local states ($M_u^* \neq M_v^*$), breaking the guarantee of value convergence.}
\label{fig:fig1}
\end{figure}
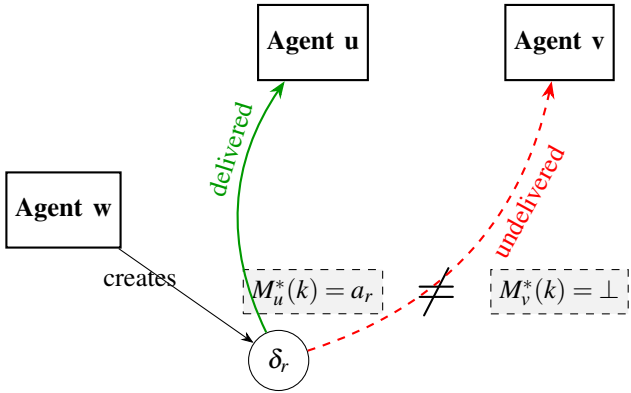

\begin{figure}[htbp]
\centering
\begin{tikzpicture}[
    contrib/.style={rectangle, draw, fill=blue!10},
    state/.style={rectangle, draw, thick, fill=yellow!20, minimum width=1.5cm}
]
    \begin{scope}[local bounding box=partA]
        \node {\textbf{(a) Schedule A: $\delta_1$ then $\delta_2$}};
        \node[state] (s0a) [below=0.5cm] {$S=0$};
        \node[state] (s1a) [below=of s0a] {$S=1$};
        \node[state] (s2a) [below=of s1a] {$S=2$};

        \draw[-{Stealth}] (s0a) -- (s1a) node[midway, right] {apply $\delta_1(val=1)$};
        \draw[-{Stealth}] (s1a) -- (s2a) node[midway, right] {apply $\delta_2(val=2)$};
        \node[right=0.2cm of s2a, red, font=\bfseries] {Final};
    \end{scope}

    \begin{scope}[yshift=-5.5cm, local bounding box=partB]
         \node {\textbf{(b) Schedule B: $\delta_2$ then $\delta_1$}};
        \node[state] (s0b) [below=0.5cm] {$S=0$};
        \node[state] (s2b) [below=of s0b] {$S=2$};
        \node[state] (s1b) [below=of s2b] {$S=1$};

        \draw[-{Stealth}] (s0b) -- (s2b) node[midway, right] {apply $\delta_2(val=2)$};
        \draw[-{Stealth}] (s2b) -- (s1b) node[midway, right] {apply $\delta_1(val=1)$};
        \node[right=0.2cm of s1b, red, font=\bfseries] {Final};
    \end{scope}
\end{tikzpicture}
\caption{Violation of Axiom~\ref{ax:semilattice} (Join-Semilattice). Without a commutative and idempotent merge operation (e.g., using an ``overwrite'' logic), the final state becomes dependent on the arbitrary message arrival order, thus violating determinism.}
\label{fig:fig2}
\end{figure}

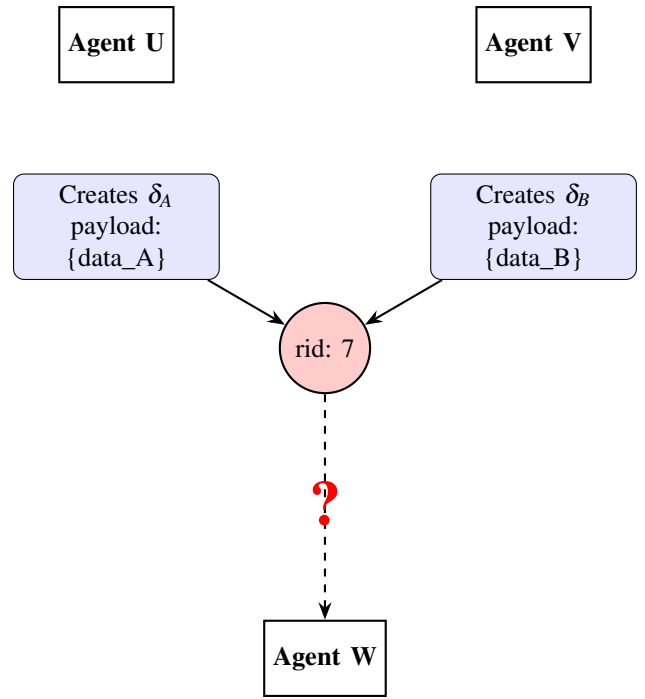
\begin{figure}[htbp]
\centering
\begin{tikzpicture}[
    node distance=1.2cm and 1.5cm, 
    agent/.style={rectangle, draw, thick, minimum size=1cm, font=\bfseries},
    contrib/.style={rectangle, draw, rounded corners, fill=blue!10, text width=2.5cm, align=center},
    rid_node/.style={circle, draw, thick, fill=red!20, minimum size=1.2cm},
    arrow/.style={-Stealth, thick}
]
    \node[agent] (u) {Agent U};
    \node[agent] (v) [right=4cm of u] {Agent V};

    \node[contrib] (delta_a) [below=of u] {Creates $\delta_A$ \\ payload: \{data\_A\}};
    \node[contrib] (delta_b) [below=of v] {Creates $\delta_B$ \\ payload: \{data\_B\}};

    \node[rid_node] (rid7) at ($(delta_a)!0.5!(delta_b)$) [below=1cm] {rid: 7};
    
    \draw[arrow] (delta_a) -- (rid7);
    \draw[arrow] (delta_b) -- (rid7);

    \node[agent] (w) [below=3cm of rid7] {Agent W};
    \draw[arrow, dashed] (rid7) -- (w);
    \node[font=\Huge\bfseries, red] at ($(rid7)!0.5!(w)$) {?};

\end{tikzpicture}
\caption{Violation of \textbf{Axiom \ref{ax:uniqueness} (Contribution as an Immutable Fact)}. If the uniqueness of a \texttt{rid} is not guaranteed, two different contributions ($\delta_A$ and $\delta_B$) can claim the same identity. This makes the global graph ill-defined and violates structural determinism.}
\label{fig:fig3}
\end{figure}

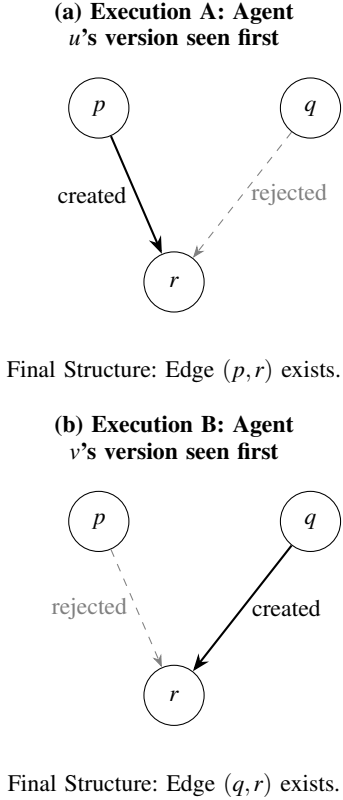
\begin{figure}[htbp]
\centering
\begin{tikzpicture}[
    font=\small,
    node distance=1cm,
    mynode/.style={circle, draw, minimum size=0.8cm},
    ghost/.style={circle, draw, dashed, text=gray}
]
    \begin{scope}[local bounding box=partA]
        \node[mynode] (pA) {$p$};
        \node[mynode] (qA) [right=2cm of pA] {$q$};
        \node[mynode] (rA) [below=1.5cm of pA, xshift=1cm] {$r$};
        
        \node[above=0.3cm of pA, xshift=1cm, text width=6cm, align=center] 
            {\textbf{(a) Execution A: Agent $u$'s version seen first}};
        
        \draw[-{Stealth}, thick] (pA) -- (rA) node[midway, left] {created};
        \draw[-{Stealth}, dashed, gray] (qA) -- (rA) node[midway, right] {rejected};
        
        \node[below=0.5cm of rA, text width=5cm, align=center, font=\small] 
            {Final Structure: Edge $(p,r)$ exists.};
    \end{scope}

    \begin{scope}[yshift=-5.5cm, local bounding box=partB]
        \node[mynode] (pB) {$p$};
        \node[mynode] (qB) [right=2cm of pB] {$q$};
        \node[mynode] (rB) [below=1.5cm of pB, xshift=1cm] {$r$};
        
        \node[above=0.3cm of pB, xshift=1cm, text width=6cm, align=center] 
            {\textbf{(b) Execution B: Agent $v$'s version seen first}};
        
        \draw[-{Stealth}, dashed, gray] (pB) -- (rB) node[midway, left] {rejected};
        \draw[-{Stealth}, thick] (qB) -- (rB) node[midway, right] {created};

        \node[below=0.5cm of rB, text width=5cm, align=center, font=\small]
            {Final Structure: Edge $(q,r)$ exists.};
    \end{scope}
\end{tikzpicture}
\caption{Violation of \textbf{Axiom \ref{ax:parents} (Immutability of Causal Linkage)}. If the \texttt{parents} metadata is mutable, two valid executions can produce non-isomorphic causal graphs. This is a direct illustration of the \textit{Structural Ambiguity} problem, which violates the guarantee of a unique DCS.}
\label{fig:fig4}
\end{figure}

\begin{figure}[htbp]
\centering
\begin{tikzpicture}[
    node distance=2cm, 
    mynode/.style={circle, draw, minimum size=0.8cm} 
]
    \node[mynode] (r1) {$r_1$};
    \node[mynode] (r2) [right=of r1] {$r_2$}; 


    \draw[-{Stealth}] (r1) to[bend left=30] node[midway, above] {\small parents=\{$r_2$\}} (r2);
    \draw[-{Stealth}] (r2) to[bend left=30] node[midway, below] {\small parents=\{$r_1$\}} (r1);
\end{tikzpicture}
\caption{Violation of Axiom~\ref{ax:causal} (Causal Well-Formedness). The prohibition of forward-references is essential. Its removal allows for the formation of cycles, breaking the foundational Directed Acyclic Graph (DAG) property required for a DCS.}
\label{fig:fig5}
\end{figure}
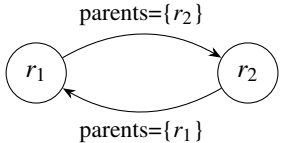

The proof of axiom minimality provides the final structural reinforcement for our theoretical edifice. It demonstrates that our proposed axiom set is precisely the necessary set of ``physical laws'' required to guarantee a DCS. At this point, we have fully defined our theory, proven its powerful internal properties, and established the rigor of its foundations. The final logical step, therefore, is to formally position this complete theory against the state-of-the-art models in the field of distributed consistency. The next proposition will, through a formal comparison, clearly reveal the essential differences and superiority of a DCS over models such as CRDTs.

\subsection{The Formal Positioning: Superiority over Value Convergence}

Finally, to precisely locate our contribution within the landscape of distributed consistency models, we formally prove that the structural guarantee of a DCS is strictly stronger than the value convergence offered by models such as CRDTs.

\begin{figure}[htbp]
\centering
\begin{tikzpicture}[
    node distance=2cm,
    mynode/.style={circle, draw, minimum size=0.8cm},
    lbl/.style={font=\small} 
]
    \begin{scope}[local bounding box=partA]
        \node[mynode] (rx) {$r_x$};
        \node[mynode] (ry) [right=of rx] {$r_y$};
        
        \node[above=0.5cm of rx, xshift=1cm] {\textbf{(a) Concurrent Execution}};
        \node[below=0.5cm of rx, xshift=1cm, lbl] {Final Value: $S = \{x\} \cup \{y\} = \{x,y\}$};
        \node[below=1.0cm of rx, xshift=1cm, lbl] {Structure: $\{r_x, r_y\}$ are concurrent};
    \end{scope}

    \begin{scope}[yshift=-4cm, local bounding box=partB] 
        \node[mynode] (ry2) {$r_y$};
        \node[mynode] (rxy) [right=of ry2] {$r_{x|y}$};
        
        \draw[-{Stealth}] (ry2) -- (rxy);
        
        \node[above=0.5cm of ry2, xshift=1cm] {\textbf{(b) Causal Execution}};
        \node[below=0.5cm of ry2, xshift=1cm, lbl] {Final Value: $S = \{y\} \cup \{x\} = \{x,y\}$};
        \node[below=1.0cm of ry2, xshift=1cm, lbl] {Structure: $r_y$ is a parent of $r_{x|y}$};
    \end{scope}
\end{tikzpicture}
\caption{Minimal construction demonstrating the separation of the DCS guarantee from CRDTs, arranged vertically for clarity. Both executions yield the same final value, satisfying CRDT guarantees. However, their causal structures are non-isomorphic,  a distinction captured by a DCS but not by value-centric models.}
\label{fig:fig6}
\end{figure}
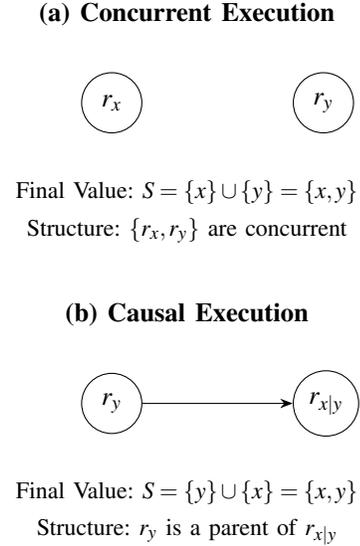

\begin{proposition}[Separation from CRDTs]\label{prop:crdt_separation}
There exists a system that satisfies the standard CRDT conditions (a join-semilattice state, an associative, commutative, and idempotent merge function, and fair communication) where the state value is guaranteed to converge, yet two valid executions can produce non-isomorphic causal histories.
\end{proposition}

\begin{IEEEproof}[Proof Sketch]
The proof relies on a simple constructive example, as illustrated in Fig.~\ref{fig:fig6}. Two contributions can be generated concurrently in one execution and sequentially (causally) in another. Both executions yield the same final state value (e.g., the set union of their payloads), satisfying CRDT guarantees. However, their underlying causal structures are fundamentally different. A DCS, through its enforcement of immutable causal metadata (Axioms~\ref{ax:uniqueness} and~\ref{ax:causal}), distinguishes between these two histories, whereas a value-centric model cannot. This demonstrates the formal separation and superiority of our structural guarantee.
\end{IEEEproof}

This final proposition concludes the formal presentation of our core theory. In this section, we have built our argument from the ground up: we first proved the existence of a Deterministic Causal Structure and revealed its critical policy-agnostic property; we then reinforced the theory's foundations by proving the minimality of its axioms; and finally, we established its novelty and superiority through a formal comparison with value-convergence models. A complete and self-consistent theoretical system for structural convergence is now established. The following sections will step back from the formal proofs to discuss the broader implications and practical considerations of this new theory.

\section{Illustrative Constructions}

The core theoretical results of this paper have been formally established in Section~IV. To aid intuition, this section presents a set of \textit{illustrative constructions} designed to visually and quantitatively demonstrate the implications of our theorems. These examples are \textbf{not performance benchmarks}, but rather serve to reinforce the correctness and necessity of our formal theory.

\subsection{Axiom Necessity}
To illustrate Theorem~\ref{thm:minimality} (Axiom Minimality), we compare a fully DCS-compliant system against variants where key integrity axioms are violated. For each case, we execute the simulation twice from an identical initial state and test for structural isomorphism in the resulting histories. A non-zero ambiguity rate indicates a failure of determinism.

\begin{table}[htbp]
\centering
\caption{Demonstration of Structural Ambiguity upon Axiom Violation}
\label{tab:illustrative_ambiguity}
\begin{tabular*}{\columnwidth}{@{\extracolsep{\fill}} l l r}
\toprule
\textbf{System Type} & \textbf{Violated Axiom(s)} & \textbf{Ambiguity Rate} \\
\midrule
DCS (Baseline) & \textit{None} & \textbf{0\%} \\
\midrule
Non-DCS Control & Metadata Mutability (Axiom 4) & 100\% \\
Non-DCS Control & Causal Forgery (Axiom 5) & 100\% \\
\bottomrule
\end{tabular*}
\end{table}

The results starkly demonstrate that removing any core integrity axiom collapses determinism into total ambiguity, visually confirming that the axiom set is indeed minimal and necessary.

\subsection{Policy-Agnosticism}
To illustrate Theorem~\ref{thm:B} (The Decoupling Invariant), we compare three heterogeneous policies: a static optimal routing strategy, a reinforcement-learning Q-routing agent, and an adaptive variant. Despite having radically different internal mechanics, for the subset of tasks where all three policies produced the exact same final path, their resulting Provenance DAGs were found to be isomorphic in \textbf{100\%} of cases. This confirms that the DCS invariant correctly captures \textit{what happened} independent of \textit{how it happened}, thereby formally decoupling correctness from policy.

\subsection{Distinctiveness of Policies}
Finally, to validate that the policies being compared are genuinely heterogeneous, we briefly present their distinct performance characteristics.

\begin{table}[htbp]
\centering
\caption{Performance Characteristics of Heterogeneous Policies}
\label{tab:illustrative_performance}
\begin{tabular*}{\columnwidth}{@{\extracolsep{\fill}} l c c r}
\toprule
\textbf{Policy} & \textbf{Mean Hops} & \textbf{Std. Dev.} & \textbf{Success (\%)} \\
\midrule
Static Optimal & 4.02 & 1.24 & 100\% \\
Q-Routing & 4.29 & 1.35 & 100\% \\
Adaptive Q-Routing & 4.38 & 1.50 & 100\% \\
\bottomrule
\end{tabular*}
\end{table}

While the causal structures they produce are identical when their execution paths align, their performance characteristics remain quantifiably different. This demonstrates that the DCS invariant preserves correctness without erasing meaningful diversity in policy efficiency.

In summary, these minimal constructions visually confirm the necessity of the axioms and the policy-agnostic nature of the DCS. All formal guarantees, however, are already proven in Section~IV; this section merely serves to illustrate those guarantees in action.

\section{Discussion and Practical Considerations}

The preceding section established the formal theory of the DCS as a policy-agnostic invariant. This section steps back from the formal proofs to explore the broader significance of this result, arguing that the \textbf{decoupling of correctness from policy} is not merely a theoretical curiosity, but a powerful new paradigm for designing and evolving complex distributed systems.

\subsection{A New Design Paradigm: \textbf{Correctness-as-a-Chassis}}

\begin{figure}[h!]
    \centering
    \includegraphics[width=0.9\linewidth]{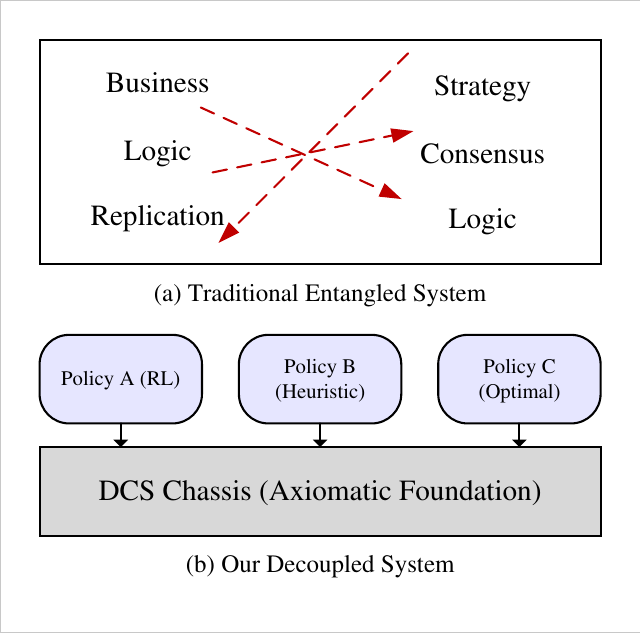}
    
    \caption{Architectural comparison between traditional systems and our DCS framework. (a) In traditional systems, policy logic is deeply entangled with correctness logic. (b) Our framework provides a \textbf{Correctness-as-a-Chassis}, where an axiomatic foundation allows diverse policies to be plugged in interchangeably without compromising structural integrity.}
    \label{fig:chassis_concept_intro}
\end{figure}

The core implication of our theory is a paradigm shift in system design. Traditionally, correctness is a holistic emerging property of the entire system, including its policies. Our work reframes correctness as a fixed, verifiable \textbf{"chassis"}, the DCS, upon which any compatible operational \textbf{"engine"}, the policy, can be mounted. This \textbf{Correctness-as-a-Chassis} model, illustrated in Fig.~\ref{fig:chassis_concept_intro}, has profound theoretical and practical implications.

\paragraph{Safe and Independent Evolution}
The single greatest benefit of this decoupling is enabling the safe and independent evolution of a system's performance and its logic. Engineering teams can aggressively optimize policies (e.g., implementing a new batching strategy for higher throughput or a speculative execution model for lower latency) without fear of corrupting the system's structural integrity. As long as the new policy respects the axioms at the contribution level, the correctness of the overall system, as captured by the DCS, is guaranteed. This dramatically reduces the complexity of system evolution and enables continuous, safe refactoring.

\paragraph{Formal Verifiability and Auditing}
The DCS transforms the system's interaction history from an opaque, policy-dependent artifact into a deterministic "white box". Because the final Provenance DAG is unique and constructible, it serves as an immutable ground truth for the system's entire causal history. This provides a solid foundation for high-stakes applications requiring formal verification, auditing, and replay, such as in decentralized finance or safety-critical autonomous systems. The audit is performed on the structure, which is guaranteed to be independent of any specific policy that was running at the time.

\paragraph{Composable Systems}
Decoupling provides a clean contract for system composition. Two or more systems, each built upon a DCS foundation, can be composed with a much higher degree of confidence. The interaction between them can itself be modeled as a set of contributions, resulting in a higher-level DCS. This allows for modular reasoning, where the correctness of each component can be analyzed independently of the others' internal policies, paving the way for a true marketplace of verifiable, interoperable intelligent components.

\subsection{System Design Implications: The Power of Decoupling}

The "Correctness-as-a-Chassis" paradigm has direct and powerful implications for system design, simplifying implementation and unlocking performance optimization.

\paragraph{Implementation Simplicity and Emergent Consensus}
Implementing the core guarantees of the DCS theory is remarkably lightweight. Because correctness is guaranteed by the immutable, axiomatic structure of the data itself, the system does not require complex, state-based consensus protocols like Paxos or Raft. A minimal implementation consists of only three simple parts: 1) a data structure for contributions with immutable metadata (\texttt{rid}, \texttt{parents}); 2) any communication protocol, such as standard gossip, that satisfies weak fairness (Axiom~\ref{ax:fairness}); and 3) a local key-value store. In such a system, consensus is not actively "negotiated" through multi-phase commits; it is an Emergent Consensus---a property that is guaranteed to arise naturally from the structure of the data itself. 

However, it must be acknowledged that while the consensus logic is lightweight, maintaining the complete Provenance DAG introduces overhead in storage and communication (e.g., the `parents` metadata). This is the necessary trade-off for achieving the stronger guarantee of structural determinism over mere value convergence. Managing this overhead through techniques like graph pruning or garbage collection is therefore a key area for engineering optimization.

\paragraph{Performance as an Independent Optimization Layer}
This minimal implementation prioritizes correctness and simplicity, not raw performance. However, the true power of the decoupling paradigm is that it treats \textbf{performance as an independent, pluggable optimization layer}. The DCS axioms provide the rigid "correctness chassis." Upon this foundation, engineers are free to design, test, and deploy a wide array of performance-enhancing policies without risking the system's structural integrity. Examples of such independent policy optimizations include:
\begin{itemize}
    \item \textbf{Intelligent Gossiping:} Designing propagation protocols that prioritize certain messages or prune redundant transmissions to speed up convergence.
    \item \textbf{Graph Sharding:} Partitioning the DCS graph across nodes to handle massive scale.
    \item \textbf{Fast Paths:} Implementing specialized, low-latency protocols for specific, non-contentious interaction patterns.
\end{itemize}
These optimization avenues, safely insulated from the core correctness logic, represent a rich and promising area for future engineering research.

\subsection{The Scope of Correctness: A Two-Layer Model}

It is crucial to precisely define the scope of "correctness" that our theory guarantees to decouple from policy. We propose a two-layer model for understanding correctness in complex distributed systems:

\begin{itemize}[leftmargin=*]
    \item \textbf{Layer 1: Causal History Correctness.} This foundational layer concerns the integrity of the system's historical record. It asks: "Did all participants agree on a single, unique, and immutable history of what events occurred and their causal relationships?" A system that fails at this layer is fundamentally non-auditable and non-deterministic.

    \item \textbf{Layer 2: Application Semantic Correctness.} This upper layer is built upon a consistent historical record and concerns the business logic. It asks: "Does the sequence of events in the agreed-upon history violate any application-specific invariants (e.g., 'a resource was not double-spent,' 'an account balance never dropped below zero')?"
\end{itemize}

The core contribution of the DCS theory is to provide a formal guarantee for \textbf{Layer 1 Correctness} and to decouple it entirely from operational policies. The DCS serves as the immutable ground truth—the "what happened"—upon which Layer 2 correctness can be verified.

Our theory does not, and cannot, automatically guarantee Layer 2 correctness, as that is the responsibility of the application logic itself. For instance, if an application policy erroneously generates two "spend" contributions for the same resource, the DCS will faithfully and deterministically record both events in the causal history. The history itself will be correct (Layer 1), but it will record a violation of application semantics (Layer 2).

By providing a solid, policy-agnostic foundation for Layer 1, the DCS dramatically simplifies the problem of reasoning about and enforcing Layer 2 correctness, as developers are freed from the complexities of asynchronous message passing and can focus solely on the logical validity of the event sequence itself.

\subsection{The Two Decouplings: Execution Path vs. Recording Mechanism}

To fully appreciate the scope of our theory, it is helpful to distinguish between two levels of system behavior, each with its own relationship to "policy":

\begin{itemize}[leftmargin=*]
    \item \textbf{The Execution Path:} This refers to the specific, ordered set of contributions that are actually generated by the agents over time. This path is inherently \textbf{policy-dependent}, as it is the direct result of the agents' high-level decision-making or "generation" policies. Different strategies will inevitably lead to different execution paths.

    \item \textbf{The Recording Mechanism:} This refers to the infrastructural process by which an execution path is observed, propagated, and solidified into a global, consistent causal history (the DCS). Our theory proves that this mechanism is \textbf{policy-agnostic} with respect to operational policies (routing, scheduling, batching).
\end{itemize}

The core decoupling achieved by our work is at the level of the Recording Mechanism. We do not eliminate the dependency of the system's outcome on agent strategy; rather, we provide a foundational guarantee that for any strategic path taken by the agents, the recording of that path is deterministic, reliable, and free from the influence of the underlying network and processing mechanics. This allows for a clean separation of concerns, enabling developers to reason about high-level agent strategy (the Execution Path) on top of a verifiably sound historical foundation (the Recording Mechanism).

\subsection{Limitations and Future Work}

While the DCS theory provides a powerful deterministic foundation for distributed intelligent systems, its current theoretical boundaries and future development trajectory must be clearly articulated.

\paragraph{Scope and Future Directions}
The scope of this work is focused on non-Byzantine environments, which is a model that accurately reflects the challenges in a vast domain of collaborative, code-driven MAS. This focus allows us to establish a foundational theory of structural determinism, separate from the orthogonal complexities of adversarial, protocol-violating behavior. Extending these deterministic guarantees to Byzantine fault-tolerant (BFT) settings is therefore a natural and high-priority direction for future research, which would adapt the DCS paradigm for open and permissionless systems.

\paragraph{Future Work: A Clear Roadmap}
Our work opens up several exciting directions for future research:
\begin{enumerate}[label=(\roman*), wide, labelindent=0pt, leftmargin=*, itemsep=0.5ex]
    \item \textbf{Byzantine-aware DCS Theory:} Extending the DCS theory to Byzantine fault-tolerant (BFT) settings is the highest priority on our roadmap. A promising technical path is to integrate cryptographic primitives with our axiom system. For instance, by requiring each contribution to be digitally signed by its creator, we can strengthen Axioms~\ref{ax:uniqueness} and~\ref{ax:causal} against malicious tampering.
    \item \textbf{Performance Evaluation of a DCS Prototype:} The second major direction is engineering implementation and performance evaluation. We plan to design and implement a prototype DCS system. Based on this prototype, we will conduct quantitative experiments on the various performance optimization strategies discussed previously (e.g., intelligent gossip, sharding, fast paths). This will provide critical performance data and engineering insights for the practical deployment of DCS.
    \item \textbf{An Application Framework atop DCS:} Ultimately, our goal is to empower application developers with the DCS theory. To this end, we plan to develop a programming framework or library built on top of the DCS theory. This framework will encapsulate the implementation details of the axioms and provide developers with a clean API, enabling them to easily build a new generation of distributed AI applications with verifiability built-in.
\end{enumerate}

\section{Conclusion}

This paper introduced the \textit{Deterministic Causal Structure (DCS)} as a theoretical foundation for decoupling correctness from policy in multi-agent systems. By axiomatizing contributions, we proved (i) \textbf{Existence and Uniqueness} (Theorem~1), (ii) \textbf{Policy-Agnostic Invariance} (Theorem~2), (iii) \textbf{Observational Equivalence} (Proposition~1), and (iv) \textbf{Axiom Minimality} (Theorem~3).  

Together these results identify DCS as a \textit{boundary principle} in distributed systems, comparable to CAP or FLP: any system that aspires to coordination-free determinism must constrain its state model to the expressive power of a join-semilattice. Correctness is thus reframed not as a protocol property, but as a structural invariant of asynchronous computation.  

To aid intuition, we included a few \textit{illustrative constructions} that visualize the implications of our theorems. These are not experiments but minimal demonstrations; all guarantees remain purely theorem-driven.  

Finally, this perspective suggests the broader design paradigm of \textbf{Correctness-as-a-Chassis}: correctness as a fixed, policy-agnostic substrate on which strategies can evolve independently. This principle opens the path toward modular and verifiably safe distributed intelligent systems, establishing DCS as a foundational building block for the mathematics of distributed intelligence.

\appendices
\section{Formal Proofs of Foundational Lemmas}
\label{sec:appendix_lemmas}

\begin{lemma}[State Monotonicity]\label{lem:monotonicity_appendix}
For any agent $i$ and key $k$, its local state sequence $\{M_i(k,t)\}_{t\in\mathbb{N}}$ is monotone non-decreasing with respect to the partial order $\sqsubseteq$ of the dcpo-join-semilattice $(L_k, \sqsubseteq, \sqcup)$.
\end{lemma}
\begin{IEEEproof}
The local state update rule is defined as $M_i(k, t+1) \leftarrow M_i(k, t) \sqcup \mathrm{payload}(\delta)$. According to Axiom~\ref{ax:semilattice}, the state space is a join-semilattice where the join operation $\sqcup$ is inflationary, meaning $x \sqsubseteq x \sqcup y$ for all elements $x, y$. Therefore, $M_i(k,t) \sqsubseteq M_i(k,t+1)$ holds for every update, establishing the monotonicity of the state sequence.
\end{IEEEproof}

\begin{lemma}[Order and Duplicate Irrelevance on Directed Sets]\label{lem:irrelevance_appendix}
For any finite or directed set of contributions for a given key $k$, the final merged state is independent of both the arrival order and the multiplicity of each contribution.
\end{lemma}
\begin{IEEEproof}
Let $\{\delta_j\}_{j\in J}$ be a finite or directed set of contributions. The final state is the join of all payloads: $\bigsqcup_{j\in J} \mathrm{payload}(\delta_j)$. By Axiom~\ref{ax:semilattice}, the join operation $\sqcup$ is associative and commutative, which guarantees order-invariance. The idempotence of $\sqcup$ ($x \sqcup x = x$) ensures that duplicate contributions do not alter the result. The directed-completeness property guarantees that this supremum exists for any directed set of contributions.
\end{IEEEproof}

\begin{lemma}[Decomposability and Traceability]\label{lem:traceability_appendix}
An agent's local state is always equal to the join of all unique contribution payloads it has observed. Furthermore, the causal ancestry of any contribution is unambiguously traceable through the immutable `parents` relation.
\end{lemma}
\begin{IEEEproof}
By construction, the local state $M_i(k)$ is formed exclusively by applying the join operation. By induction and Lemma~\ref{lem:irrelevance_appendix}, $M_i(k) = \bigsqcup_{\delta \in S_i(k)} \mathrm{payload}(\delta)$, where $S_i(k)$ is the set of unique contributions received by agent $i$ for key $k$. The traceability of causal history is a direct consequence of Axioms~\ref{ax:uniqueness} and~\ref{ax:parents}, which state that the `rid` and `parents` set of each contribution are immutable and globally consistent. This fixes the vertices and edges of the provenance graph at the moment of creation.
\end{IEEEproof}

\begin{lemma}[Eventual Propagation]\label{lem:propagation_appendix}
For any key $k$, every contribution $\delta$ created for that key will eventually be delivered at least once to every agent in the relevant set $\mathrm{Rel}(k)$.
\end{lemma}
\begin{IEEEproof}
This is a direct restatement of Axiom~\ref{ax:fairness} (Localized Weak Fairness). The axiom guarantees that if a contribution is persistently available in the network, all interested agents will eventually receive it. By Lemma~\ref{lem:irrelevance_appendix}, subsequent duplicate deliveries are harmless.
\end{IEEEproof}

\begin{lemma}[Information Preservation]\label{lem:preservation_appendix}
Once a payload has been incorporated into an agent's local state via the join operation, its information is never lost or overwritten by subsequent updates.
\end{lemma}
\begin{IEEEproof}
This follows directly from the state monotonicity established in Lemma~\ref{lem:monotonicity_appendix}. Since any subsequent state $M'$ is computed as $M' = M \sqcup x$, it is guaranteed that $M \sqsubseteq M'$. Thus, the information contained in the prior state $M$ is preserved as a lower bound of all future states.
\end{IEEEproof}

\section{Proofs of Core Theorems}
\label{sec:appendix_theorems}

\begin{IEEEproof}[Proof of Theorem~\ref{thm:A}]
We prove the three claims separately.
\paragraph{1. Existence} The vertices of the graph $G^*$ are the set of all contributions $\mathcal{R}$. By Axiom~\ref{ax:uniqueness}, each contribution has a unique identifier (`rid`). The edges are defined by the immutable `parents` metadata. Axiom~\ref{ax:causal} (Causal Well-Formedness) mandates that an agent can only list `rid`s of contributions it has already observed in the `parents` set of a new contribution. This enforces a strict temporal ordering on edge creation, making it impossible to form cycles. Therefore, $G^*$ is a well-defined Directed Acyclic Graph (DAG).

\paragraph{2. Uniqueness} The structure of $G^*$ is determined solely by its vertices (`rid`s) and edges (`parents` links). Axioms~\ref{ax:uniqueness} and~\ref{ax:parents} guarantee that this metadata is immutable upon creation. Agent policies—such as scheduling, batching, or routing—only affect the *timing* and *path* of information propagation, not the content of the immutable metadata itself. Since the graph's definition is independent of any such policy, its structure is unique up to isomorphism for any given set of contributions.

\paragraph{3. Constructibility} For any agent $i \in \mathrm{Rel}(k)$, its local state sequence $\{M_i(k,t)\}$ forms a monotone chain (Lemma~\ref{lem:monotonicity_appendix}) in a directed-complete partial order (Axiom~\ref{ax:semilattice}). This guarantees the existence of a limit state (the supremum). By Lemma~\ref{lem:propagation_appendix}, agent $i$ will eventually receive every contribution. By Lemma~\ref{lem:irrelevance_appendix}, the final joined state is independent of arrival order. Thus, the limit state is unique and equal to the join of all payloads in the final, globally consistent mergeable set.
\end{IEEEproof}

\begin{IEEEproof}[Proof of Theorem~\ref{thm:B}]
This theorem is a direct consequence of Theorem~\ref{thm:A}. The proof of uniqueness in Theorem~\ref{thm:A} already established that the graph's structure is determined by immutable metadata, which is independent of agent policy. Therefore, if the set of contributions is the same, the resulting graphs $G_1^*$ and $G_2^*$ must be isomorphic. The convergence to identical local states follows from the constructibility proof, which showed the limit state is the join over the complete set of payloads, a calculation that is itself order-independent and thus policy-agnostic.
\end{IEEEproof}

\section{Proofs of Supporting Propositions}
\label{sec:appendix_propositions}

\begin{IEEEproof}[Proof of Proposition~\ref{prop:C}]
For computations that only use causal queries (ancestor/descendant) and semilattice-homomorphic aggregations, two executions are observationally indistinguishable if and only if their DCS graphs are isomorphic.

($\Leftarrow$) If the DCS graphs $G_1^*$ and $G_2^*$ are isomorphic, there exists a bijection that preserves \texttt{rid}s, \texttt{parents} relations, and \texttt{payload}s. Any causal query on $G_1^*$ will yield a result that maps directly to the identical query on $G_2^*$. Any aggregation over payloads will operate on identical sets of data and, due to the properties of the join-semilattice, produce identical results. Thus, the executions are observationally indistinguishable.

($\Rightarrow$) If the graphs are non-isomorphic, there must exist a structural difference. For example, a contribution $r$ in $G_1^*$ has a parent $p$ that is not a parent of $r$ in $G_2^*$. A simple query such as "Is $p$ an ancestor of $r$?" would return true in the first execution and false in the second. This constitutes a distinguishable observation. Therefore, observational indistinguishability implies graph isomorphism.
\end{IEEEproof}

\begin{IEEEproof}[Proof of Proposition~\ref{prop:crdt_separation}]
There exist systems satisfying standard CRDT conditions that converge to the same state value but produce non-isomorphic causal histories.

We prove by construction. Let the state be a set and the merge operation be set union ($\cup$), a valid join-semilattice. Consider two executions generating contributions with payloads $\{x\}$ and $\{y\}$.

\textbf{Execution 1 (Concurrent):} Agent A creates contribution $\delta_x$ (payload $\{x\}$), and Agent B concurrently creates $\delta_y$ (payload $\{y\}$). Neither is a parent of the other.
\begin{itemize}
    \item \textbf{Final Value:} All agents eventually receive both and compute the final state $\{x\} \cup \{y\} = \{x, y\}$.
    \item \textbf{Causal Structure:} The DCS graph consists of two disconnected nodes, $\{\delta_x, \delta_y\}$, representing concurrent events.
\end{itemize}

\textbf{Execution 2 (Causal):} Agent A creates $\delta_x$. Agent B observes $\delta_x$ and then creates $\delta_y$, explicitly setting $\mathrm{parents}(\delta_y) = \{\mathrm{rid}(\delta_x)\}$.
\begin{itemize}
    \item \textbf{Final Value:} All agents eventually receive both and compute the final state $\{x\} \cup \{y\} = \{x, y\}$.
    \item \textbf{Causal Structure:} The DCS graph is a two-node chain, $\delta_x \to \delta_y$, representing a causal dependency.
\end{itemize}

Both executions satisfy CRDT value convergence, arriving at the identical state $\{x, y\}$. However, their DCS graphs are non-isomorphic. This demonstrates that the structural guarantee of a DCS is strictly stronger than the value convergence guarantee of CRDTs.
\end{IEEEproof}

\section{Proof of Theorem~\ref{thm:minimality} (Axiom Minimality)}
\label{sec:appendix_minimality}

\begin{IEEEproof}[Proof of Theorem~\ref{thm:minimality}]
The set of Axioms \ref{ax:fairness} through \ref{ax:causal} is minimal. The removal of any single axiom permits a counterexample where the guarantees of Theorem~\ref{thm:A} fail.

Sufficiency was established by the proof of Theorem~\ref{thm:A}. We prove necessity by constructing a counterexample for the removal of each axiom.

\begin{description}
    \item[Without Axiom~\ref{ax:fairness} (Weak Fairness)] An agent $i \in \mathrm{Rel}(k)$ might never receive a contribution $\delta$ that other agents have received. Its local state will converge to a different limit than others, violating the \textbf{constructibility} guarantee of a unique, globally consistent state.

    \item[Without Axiom~\ref{ax:semilattice} (Join-Semilattice)] If the merge operation is not associative, commutative, and idempotent (e.g., "last-writer-wins"), the final state becomes dependent on the arbitrary network delivery order. This violates the \textbf{uniqueness} and \textbf{constructibility} guarantees, as different policies lead to different outcomes.

    \item[Without Axiom~\ref{ax:uniqueness} (RID Uniqueness and Immutability)] If two contributions could share the same `rid`, or if a contribution's `payload` could be mutated after creation, the set of vertices in the global graph $G^*$ would be ill-defined and non-static. This violates the \textbf{existence} of a single, well-defined graph.

    \item[Without Axiom~\ref{ax:parents} (Parent Set Immutability)] If the `parents` metadata were mutable, an agent could alter the causal history after the fact. Two different agents could observe and record different parent sets for the same contribution, leading to non-isomorphic graphs. This violates the \textbf{uniqueness} guarantee.
    
    \item[Without Axiom~\ref{ax:causal} (Causal Well-Formedness)] An agent could create a contribution $\delta_1$ that names a not-yet-created contribution $\delta_2$ as its parent, while the creator of $\delta_2$ names $\delta_1$ as its parent. This would create a cycle ($\delta_1 \to \delta_2 \to \delta_1$). A graph with cycles is not a DAG, violating the fundamental \textbf{existence} guarantee of the DCS.
\end{description}
Since removing any axiom breaks at least one core guarantee, the axiom set is minimal.
\end{IEEEproof}

\bibliographystyle{IEEEtran}
\bibliography{IEEEabrv,bibliography}

\end{document}